\documentclass[10pt]{article}
\usepackage{bm}
\usepackage{amsmath}
\usepackage{amsfonts,amsbsy}
\usepackage{amssymb}
\usepackage{graphicx}
\usepackage[colorlinks=true, pdfstartview=FitV, linkcolor=red, citecolor=blue, urlcolor=blue]{hyperref}

\def\empile#1\over#2{\mathrel{\mathop{\kern 0pt#1}\limits_{#2}}}

\def\beq{\begin{equation}}
\def\eeq{\end{equation}}
\def\bea{\begin{eqnarray}}
\def\eea{\end{eqnarray}}

\def\p{{\boldsymbol p}}

\def\d3p{\frac{d^3\p}{(2\pi)^3}E_\p}

\catcode`\@=11

\newcount\@tempcntc
\def\@citex[#1]#2{\if@filesw\immediate\write\@auxout{\string\citation{#2}}\fi
  \@tempcnta\z@\@tempcntb\m@ne\def\@citea{}\@cite{%
        \@for\@citeb:=#2\do%
    {\@ifundefined{b@\@citeb}%
        {\@citeo\@tempcntb\m@ne\@citea%
                \def\@citea{,\penalty\@m\ }{\bf ?}\@warning%
                {Citation `\@citeb' on page \thepage \space undefined}}%
        {\setbox\z@\hbox{\global\@tempcntc0\csname b@\@citeb\endcsname\relax}
     \ifnum\@tempcntc=\z@ \@citeo\@tempcntb\m@ne%
       \@citea\def\@citea{,\penalty\@m}%
       \hbox{\csname b@\@citeb\endcsname}%
     \else%
      \advance\@tempcntb\@ne%
      \ifnum\@tempcntb=\@tempcntc%
      \else\advance\@tempcntb\m@ne\@citeo%
      \@tempcnta\@tempcntc\@tempcntb\@tempcntc\fi\fi}}\@citeo}{#1}}%

\def\@citeo{\ifnum\@tempcnta>\@tempcntb\else\@citea
  \def\@citea{,\penalty\@m}%
  \ifnum\@tempcnta=\@tempcntb\the\@tempcnta\else
   {\advance\@tempcnta\@ne\ifnum\@tempcnta=\@tempcntb \else
\def\@citea{--}\fi
    \advance\@tempcnta\m@ne\the\@tempcnta\@citea\the\@tempcntb}\fi\fi}

\catcode`\@=12

\begin{document}

\title{\bf Electroweak Instantons, Axions, and the Cosmological Constant }
\author{Larry McLerran$^{(1,2)}$, Robert D. Pisarski$^{(1,2)}$, and Vladimir Skokov$^{(1)}$}

\maketitle

\begin{enumerate}

\item Nuclear Theory, Department of Physics, 
Brookhaven National Laboratory,
   Upton, NY 11973, USA
 \item RIKEN BNL Research Center, 
 Brookhaven National Laboratory,
   Upton, NY 11973, USA
\end{enumerate}

\begin{abstract}
If there is explicit violation of baryon plus lepton number 
at some energy scale,
then the electroweak theory depends upon a $\theta$-angle.
Due to a singular integration over
small scale size instantons, this $\theta$-dependence 
is sensitive to very high momentum scales.
Assuming that 
there is no new physics between the electroweak and Planck scales,
for an electroweak axion the energy difference between
the vacuum at $\theta \neq 0$, and that at $\theta = 0$,
is of the correct order of magnitude to be the dark energy 
observed in the present epoch. 
\end{abstract}

\section{Introduction}

In  a gauge theory, a $\theta$-angle appears by adding a term
\begin{equation}
\theta  \; {\alpha \over {8\pi}} \int~d^4x~ 
{\rm tr} \left(F_{\mu \nu} \widetilde{F}^{\mu \nu} \right)
\label{theta}
\end{equation}
to the action, where $F_{\mu \nu}$ is the field strength, and
$\widetilde{F}_{\mu \nu} 
= {1 \over 2} \epsilon_{\mu \nu \lambda \sigma} F^{\lambda \sigma}$ 
its dual.  Here $N$,
\begin{equation}
N = {\alpha \over {8\pi}}  \int~d^4x~ 
{\rm tr} \left(F_{\mu \nu} \tilde{F}^{\mu \nu} \right)
\end{equation}
is the winding number of an Euclidean field 
configuration~\cite{'tHooft:1976fv,Callan:1976je,Coleman:1978ae,Krasnikov:1978dg,'tHooft:1986nc}.   

Instantons are solutions of the field equation with finite action
and nonzero topological charge, $N \neq 0$.  
For the $SU(3)$ color gauge field, the physics of such configurations
are well known.
One can add a term such as Eq. (\ref{theta}) to the action, which violates
CP symmetry.  A small value of $\theta$ can be attained by using
the Peccei-Quinn symmetry \cite{Peccei:1977hh}, 
and coupling to an axion field \cite{Wilczek:1977pj}.

In the electroweak theory, for the $SU(2)_L$ gauge field, the
associated $\theta$ angle has no physical significance.
Since the electroweak action conserves baryon number, 
it does not change under a rotation of $(B+L)$ number, 
where $B$ and $L$ denote baryon and lepton number, respectively.
The only place where electroweak instantons contribute are in amplitudes 
connecting states with different numbers of 
baryons~\cite{'tHooft:1976fv,Krasnikov:1978dg}.
For such processes, if $(B+L)$ changes by
by $\Delta(B+L) = 3N$, then $\theta$ appears in the path integral
as ${\rm e}^{i N \theta}$.  
The factor of three arises here because each generation of quark is produced.
The basic instanton process therefore involves 
9 colored quarks and three leptons.  
In amplitudes squared,
the phase disappears and the $\theta$-angle is no consequence.

\section{Beyond the Standard Model}

In generalizations of the standard model, 
one can have processes that violate $(B+L)$ explicitly. Following Anselm 
and Johansen~\cite{Anselm:1993uj}, 
let us assume there is an explicit $(B+L)$ violating interaction of the form
\begin{equation}
 S_{(B+L)} = {1 \over M^2}~
\int~ d^4x \left\{ \lambda \; l_L \, q_L \, q_L \, q_L  + c.c. \right\} . 
\label{BLviolation}
\end{equation}
Here $l_L$ is a left handed lepton field 
and $q_L$ is a left handed quark field.  
The scale $M$ is the energy scale at which
lepton and baryon number changing interactions are important, 
and is presumably a scale of a Grand Unified Theory (GUT) or higher.  
The matrix $\lambda$ is of order $1$, 
and contracts various spinor, color and flavor indices.  
This interaction violates both $(B+L)$ and chirality.

\begin{figure}[htbp]
\begin{center}
\resizebox*{!}{6cm}{\includegraphics{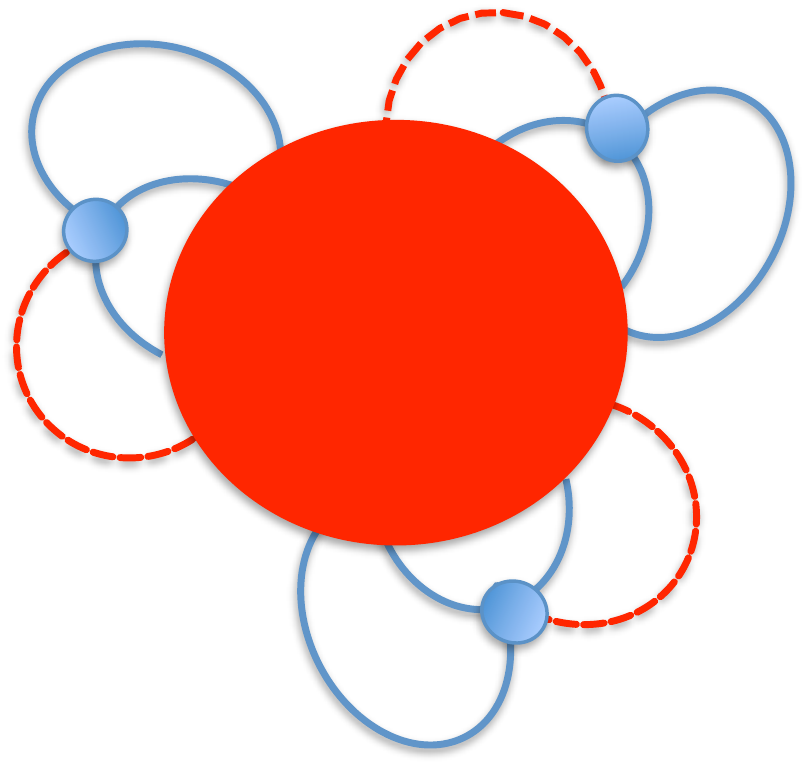}}
\end{center}
\caption{The instanton decays into 9 quarks (solid blue lines) 
and three leptons (dashed red line).  
They annihilate at the vertex associated with explicit $(B+L)$ 
violation and chirality violation. }
\label{instanton}
\end{figure}

The basic process that can generate vacuum to vacuum overlap 
is shown in Fig.~(\ref{instanton}).
A $SU(2)_L$ instanton emits three baryons and one lepton per generation;
this is compensated by vertices for the $(B+L)$ process of
Eq. (\ref{BLviolation}), so that in all,
the total amplitude does not change $(B+L)$.

One might expect that since the scale of explicit $(B+L)$ 
violation in Eq. (\ref{BLviolation})
is much larger than the electroweak scale,
that surely integrating over the instanton scale size cuts off such
efforts.  This is wrong.
After integrating over the external lines of quarks and leptons, 
the instanton amplitude is
\begin{equation}
I = \int d^4x \int {{ d\rho} \over {\rho^5}}  \; {1 \over {\rho^6 M^6}} \;
{\rm e}^{-2\pi /\alpha(\rho)}. 
\end{equation}
For the electroweak theory, the coupling constant
decreases somewhat at high momentum, or small $\rho$,
\begin{equation}
\frac{1}{\alpha(\rho)} = 
\frac{1}{\alpha(\rho_0)} 
+{{(22/3-N_f/3 - N_h/6)} \over {2\pi}} \; 
\ln(\rho_0/\rho) ,
\end{equation}
where $N_h$ is the number of Higgs particles and 
$N_f$ is the number of electroweak doublets.  
Using the values for the standard model, $N_f = 12$ and $N_h = 1$, 
we obtain $22/3-N_f/3-N_h/6 = 19/6$.  
This means that for small instantons, $\rho \gg 1/M$, the
integration over instanton scale size behaves as
\begin{equation}
\sim \frac{1}{M^6} \int {d \rho \over {\rho^{\, 47/6}}} \; .
\end{equation}
This integral does not converge, even for 
very small scale instantons above the scale $M$.
Presumably above this scale,
new physics enters which makes the $\rho$ integral convergent.

This lack of sensitivity to the low energy integration 
also occurs in supersymmetric GUT's \cite{Nomura:2000yk,Takahashi:2005kp}.  
In such theories, the electroweak coupling actually increases until 
the unification scale, with the $\rho$ integral convergent
only above the GUT scale.  Thus in supersymmetric GUT's, the
integration over instanton scale size is less convergent than without
supersymmetry.

We conclude that whatever physics is associated 
with a $SU(2)_L$
theta angle and vacuum to vacuum transitions in the electroweak theory 
is sensitive to physics at super high energy scales.

Let us assume that the scale of explicit chiral $(B+L)$ violation 
is the scale at which new physics makes the integral over
$\rho$ converge.  This scale is $M$.  
It could be a scale somewhat less than that of the Planck scale, 
where Grand Unification might occur, or it might be the Planck scale itself.  
Including the effects of zero modes \cite{Anselm:1993uj},
the rate for instanton processes is
\begin{equation}
S_{\rm I} = \kappa \; \left(\frac{2 \pi}{\alpha_{ \rm W}}\right)^4 \; 
{\rm e}^{-2\pi /\alpha(M)} \; M^4 \; ,
\end{equation}
where $\kappa$ is a constant of order 1.
If we use electroweak theory with no Grand Unification,
this formula becomes
\begin{equation}
S_{\rm I}  = \kappa \; \left(\frac{2 \pi}{\alpha_{ \rm W}}\right)^4 \; 
\left( {M_{\rm EW} \over M} \right)^{19/6}  \; 
e^{-2\pi/\alpha_{\rm W}(M_{\rm EW})} \; M^4 \; .
\end{equation}
If we take the energy scale $M$ to be the Planck mass, and 
$1/\alpha_{\rm W} \sim 1/29$, we find that
\begin{equation}
S_{\rm I} \sim 10^{-122} \cdot M_{\rm pl}^4. 
\label{darkenergy}
\end{equation}
This is remarkably close to the value of dark energy 
presently observed cosmologically, 
$\epsilon_{\rm DE} \sim 10^{-123} \cdot M_{\rm pl}^4 $
\cite{Peebles:2002gy}.
In Eq. (\ref{darkenergy}),
$M_{\rm pl} \sim 10^{19}$~GeV is the Planck mass, and not the reduced
mass, $M_{\rm pl}/\sqrt{8 \pi}$.

Note that in Grand Unified theories,
the coupling decreases less rapidly, or increases at higher energies,
making the exponential suppression less important. 
At the Planck scale, one expects that this result is 
larger than the vacuum energy, although one might argue that in a GUT, 
one would generically go to a scale lower in energy.

An estimate similar to ours has been performed
in a supersymmetric theory 
by Nomura, Watari, and Yanagida~\cite{Nomura:2000yk}, 
see also Ref.~\cite{Takahashi:2005kp}. 
In supersymmetric GUT theories, the electroweak coupling increases 
more rapidly than without supersymmetry.  Consequently, to obtain an
energy density as above, it is necessary to have a low value of the GUT
scale \cite{Nomura:2000yk,Takahashi:2005kp}.
The purpose of this note is to show that the numbers
in a model {\it without}
supersymmetry are extremely interesting in their own right. 

Tegmark, Aguirre, Rees, and Wilczek, and
later Hertzberg, Tegmark, and Wilczek \cite{Tegmark:2005dy},
consider axion models with an axion scale on order of the Planck mass.
They also find 
a reasonable value for the dark energy from the axion potential.
Their model does not involve electroweak axions, though,
and so differ in some important details from ours.

There are numerical factors that need to be computed, in order
to get a more precise estimate 
of instanton induced processes within the electroweak theory.  
There is a coefficient associated with the precise normalization of zero 
modes in the one loop computation \cite{Anselm:1993uj}.  
We have checked that using the running of $\alpha_W$ to two loop order
changes the exponent in Eq. (\ref{darkenergy}) by $\sim 1\%$.
Given all the uncertainty, the estimate we have obtained is 
remarkably close to the observed value of the dark energy.

It is interesting to consider how our estimate changes by altering the
matter of the theory.  Adding a single Higgs field has relatively little
effect, $\sim 10^{- 3}$.  In contrast, a fourth generation contributes
three quarks and one lepton, 
and so suppresses the estimate in Eq. (\ref{darkenergy})
by a signficant factor, $\sim 10^{-21}$.

One might worry that even though the 
instanton amplitude contributes to the energy of the $\theta$-vacuum 
ground state, that the rate for any such process,
which involves square of the amplitude, is so 
small that it never happens in the lifetime of the universe.  
In the evolution
of the universe from its initial conditions, though,
it should be suffice that there be an overlap between the initial set of states
and a $\theta$-vacua.
Since the lifetime of the universe, 
times the splitting of energy between the states
include instantons and those which do not include instantons,
is large, presumably one projects onto the proper ground state. 
If so, electroweak
instantons to contribute to the ground state energy as above.

\section{Electroweak Axions}

If we promote the electroweak $\theta$-angle to an axion
\cite{Peccei:1977hh,Wilczek:1977pj}
we generate a vacuum energy of order $S_I$.
Let us  take the axion field modulus, $F_A \sim M$, 
as is natural if there is one high energy scale. 
If $M$ is the Planck or a GUT scale, 
a very small axion mass is generated.
Such a light axion does not affect the long range gravitational force.
The electroweak axion is a pseudo-scalar, so that due to derivative couplings, 
at large distances the exchange of the electroweak axion
is suppressed by two powers of $r$, relative to the $1/r$ potential
of gravity.  

In the absence of Grand Unification,
it is natural to assume that $M \sim M_{\rm pl}$
\cite{Shaposhnikov:2009pv,Shaposhnikov:2010zz},
and the electroweak coupling runs up to the 
Planck scale without substantial modification.
In this case, 
the Compton wavelength for the electroweak axion is 
larger than the size of the universe.  This is automatic
since  $S_{\rm I} \sim \epsilon_{\rm vac}$, 
and $M_A \sim 1/R_{\rm universe}$ by Einstein's equations.  
For such a large value of $F_A$, the electroweak axion is
very weakly coupled, and the lower bound on $F_A$, from the cooling of 
large stars by axion emission \cite{Raffelt:1990yz}, is not a problem.
There may be additional constraints on $F_A$ where axion
emission is competitive with gravitational radiation.

Because black holes have no
hair, global symmetries can be broken by quantum effects at the Planck
scale \cite{Kamionkowski:1992mf}.
If we take a unification scale at the Planck scale, our model
is sensitive to this effect.  Having such a light axion mass may allow us
to evade this criticism; in any case, we ignore it.

The argument that the energy density trapped in 
electroweak axions is the dark energy has a naturalness problem.  
There are other vacuum energy effects that are much larger, and 
must be artificially set to zero.  
We apply a naturalness condition:
\begin{itemize}
\item After the axion field has had time to relax to zero, 
the cosmological constant should vanish. 
Therefore at late times, the
universe has vanishing cosmological constant.
\end{itemize}
If this condition is true, then 
the electroweak axion energy is non zero 
only because the modulus of the axion field has not yet relaxed to zero.  
Recall that the vacuum energy is
\begin{equation}
\epsilon_{A} \sim \; M_A^2 \; a^2 
\sim  \; M_A^2 \; F_A^2 \; \theta^2 \; \sim S_{\rm I} \; \theta^2.
\end{equation}
Because the inverse axion mass is of the order of the present size of 
the universe, there has not been time for the axion field to relax to zero.  
This happens only at some very late time, 
where the axion energy becomes dynamical,
and is ultimately a type of matter energy.
Assuming that far in the future there is no cosmological constant,
in the present epoch 
the vacuum energy in the electroweak 
axion field has physical significance.
That is, the cosmological constant is a temporary phenomenon 
that awaits its ultimate decay.

\section{Summary}

Of course the picture 
we paint of the electroweak axion as a source of the cosmological 
constant is very speculative.
Electroweak theory could have other intermediate 
energy scales that are important.  
There could be compensations and tuning of 
various energy scales so that within such a picture, 
one would get the correct dark energy.

It may also be that there is no electroweak axion, or 
that even if there is, the source of dark energy is not in the axion field.

Perhaps most interesting in these considerations is that instantons 
can be sensitive to physics in the far ultraviolet, as is also
seen in the supersymmetric case \cite{Nomura:2000yk,Takahashi:2005kp}.
There is no  
decoupling of high energy and low energy degrees of freedom. This, 
and the sensitivity of instantons to scale breaking effects suggests 
their importance in other contexts.  Perhaps in technicolor 
theories where the running of the coupling constant is assumed 
to be quite slow, effects of broken symmetries 
on higher energy scale 
might be important~\cite{Dimopoulos:1979es,Appelquist:1986an}.

It is also quite amusing that there is another indication that there may 
be no intermediate energy scale physics 
between the electroweak scale and that of the Planck scale.  
This comes from requiring that electroweak theory is 
sensible at intermediate scales, and that there are fixed points 
of coupling constant evolution at the Planck scale.  
These considerations lead to a  prediction 
of the Higgs boson mass at 126~GeV~\cite{Shaposhnikov:2009pv}.  
Such a value is suggested 
by recent results from the LHC~\cite{Aad:2012vn,Chatrchyan:2012tx}.
Schaposhnikov has argued that including massive right handed singlet
neutrinos gives a reasonable cosmology for the standard model,
with appropriate values for dark matter and the baryon
asymmetry~\cite{Shaposhnikov:2009zz}.

Lastly, we note that Zhitnitsky
has argued that there may be other processes, related to intantons
in strongly interactions, that might be responsible for the 
cosmological constant~\cite{Urban:2009yg}.

\section*{Acknowledgements}
We thank Eric Zhitnitksy for discussions and seminars
where he argued, with other mechanisms, that instantons 
in QCD might be responsible for the cosmological constant.  
We also thank Fodor Bezrukov, Hooman Davoudiasl, and Bill Marciano 
for discussions.
The research of L. McLerran, R. D. Pisarski, 
and V. Skokov is supported under DOE Contract No. DE-AC02-98CH10886.

\end{document}